\documentclass{chi2009}
\usepackage{times}
\usepackage{url}
\usepackage{graphics}
\usepackage{color}
\usepackage[pdftex]{hyperref}
\hypersetup{%
pdftitle={Gaze and Gestures in Telepresence: Multimodality, embodiment and roles
of collaboration}, pdfauthor={Mauro Cherubini, Rodrigo de Oliveira, Nuria
Oliver, and Christian Ferran}, pdfkeywords={deixis, eye-tracking, focus of
attention, gesture interfaces, natural interaction}, bookmarksnumbered,
pdfstartview={FitH}, colorlinks, citecolor=black, filecolor=black,
linkcolor=black, urlcolor=black, breaklinks=true, }
\newcommand{\comment}[1]{}
\definecolor{Orange}{rgb}{1,0.5,0}

\begin{document}

\setlength{\paperheight}{11in}
\setlength{\paperwidth}{8.5in}
\setlength{\pdfpageheight}{\paperheight}
\setlength{\pdfpagewidth}{\paperwidth}

\toappear{International Workshop New Frontiers in Telepresence 2010,\\
part of CSCW2010, Savannah, GA, USA, 7th of February, 2010.\\
\scriptsize{\url{http://research.microsoft.com/en-us/events/nft2010/}}}

\title{Gaze and Gestures in Telepresence:
multimodality, embodiment, and roles of collaboration}
\numberofauthors{1}
\author{
  \alignauthor Mauro Cherubini, Rodrigo de Oliveira, Nuria Oliver, and Christian Ferran\\
    \affaddr{Telefonica Research}\\
    \affaddr{via Augusta, 177 -- ES-08021 Barcelona, Spain}\\
    \email{\{mauro, oliveira, nuriao, icfb\}@tid.es}
}

\maketitle

\begin{abstract}
  This paper proposes a controlled experiment to further investigate
  the usefulness of \textit{gaze awareness} and \textit{gesture recognition} in the support
  of collaborative work at a distance. We propose to redesign experiments
  conducted several years ago with more recent technology that would:
   a) enable to better study of the integration of communication modalities,
  b) allow users to freely move while collaborating at a distance and c) avoid asymmetries of communication between collaborators.
\end{abstract}

\vspace{-10pt}


\section{Introduction}

Despite many years of research on Media Spaces, we are still far from developing
technologies that would allow people to collaborate at distance with the same
efficiency and ease than when face-to-face. Ethnographical observations from
real work settings show that many solutions developed to support collaborative
work at a distance are flawed as they \textit{``fracture the relation between
action and the relevant environment''} \cite{Luff:2003ij}. For example, using
many video cameras to capture and share different points of view between two
remote locations might seem to be an improvement over the use of a single
camera. However, users might feel lost in the attempt to understand which view
is the partner currently looking at or how to adapt common communication
strategies to this multitude of perspectives. As Luff {\em et al.}~explain
\cite[p.~73]{Luff:2003ij}: \textit{``Ironically, the more we attempt to enhance
the environment, the more we may exacerbate difficulties for the participants
themselves in the production and coordination of action''}. 

Similarly, our argument is that we need to find more subtle technological
solutions to translate communication mechanisms which are effective in presence
but not available when collaborators are not co-located. These solutions should
allow to recreate the same functions using different but equivalent strategies.
In this position paper we focus on two of these mechanisms, namely the awareness
of the \textit{focus of attention} and the use of \textit{gestures},
particularly deictic gestures, to disambiguate references used during the
interaction (\emph{e.g.,} discussing blood test reports from different patients)
or to better support comparisons between various information media (\emph{e.g.,}
combining a broken leg x-ray result with a plastic leg miniature so that the
physician can point specific articulations in the former and manipulate the
latter while explaining the injury cause).

We concentrate on these two elements because linguistic theories have shown
that when we communicate the production of our elocution is inextricably linked
to the responses of our audience. It is crucial for the speaker to monitor
his/her audience for evidence of continued attention and understanding
\cite{clark1991}. Clark \cite{Clark:2003qy} explained that
communication is ordinarily anchored to the material world and that one way
it gets anchored is through \textit{pointing}. Pointing-to can be achieved through
the linguistic channel using specific terms like ``this'', ``he'', or ``here''. Also,
deixis can be produced through gestures. Furthermore, it has been suggested that
deixis is intertwined with gaze awareness because face-to-face gestures can be
perceived and acknowledged by recipients using gaze \cite{Clark:2004fk}.

Research in this area has been extremely active in these last few years.
As we will detail in the next section, scholars have designed and tested interfaces
that support gaze awareness (\textit{e.g.}, \cite{Monk:2002lr, Qvarfordt:2005})
and gesturing (\textit{e.g.}, \cite{gutwin2002, Ishii:1992fk, Kirk:2007fk}, etc.).
However, we believe that there are three important
aspects that still require further consideration and research.
First, researchers should develop a careful \textit{(a) integration between
the communication modalities} because human communication is intrinsically multimodal.
When we are face-to-face we tend to use both linguistic and non-linguistic channels
to minimize the communicative effort and maximize the outcome of the interaction.
Little research so far has compared different solutions to combine communication
modes, and few studies have focused on the effect of different combinations
on collaboration.

Second, designers should enable collaboration environments that \textit{(b) ease
users transition between digital and physical workspaces}. When we are
co-located, we tend to use seamlessly (\textit{i.e.}, embody) the space around
us. We can point to digital artifacts on the screens of our computational
devices and at the same time to physical objects located nearby. However, many
telepresence prototypes designed in the past have been somewhat limited in
letting users transition from one modality to the other. Therefore, future work
should focus in enhancing the capabilities of telepresence environments so to
enable easy transition from digital environments to physical workspaces.

Finally, researchers should embrace a more \textit{(c) flexible definition of
roles within the context of collaboration}. Many environments for telepresence
studied in the past were built following a helper-worker scenario where a remote
expert could provide instructions to an on-site worker (\textit{e.g.},
\cite{Kirk:2007fk, Kuzuoka:1994wq, Monk:2002lr}). Although motivated by real
situations of the use of communication technology, telepresence prototypes that
assign static roles within the collaboration are unrealistic and create
communication asymmetries that are generally non existing in face-to-face
scenarios.

We will expand these three points in the next section. Subsequently, we will
introduce a research framework that could be used to conduct further research
on the above issues and we will discuss the expected outcomes of this research.

\section{Collaboration in Dual Spaces}

Effective collaboration requires participants to be able to communicate
their intents, agree on a methodology to achieve their goals, share
information and monitor the development of their interaction. Daly-Jones
{\em et al.}~\cite{Daly-Jones:1998bs} defined four pragmatic needs that must be
fulfilled in human interaction: 1) the need to make contact; 2) the need to
allocate turns for talking; 3) the need to monitor understanding and audience
attention; and finally 4) the need to support deixis. The last two points are
particularly interesting for the design of systems for remote collaboration, as
we will detail in the next subsections.

\subsection{Gaze and the Focus of Attention}
Previous research has demonstrated how gaze is connected to attention and, in turn,
to cognition \cite{}. Gaze is also used to marshal turn-taking. Therefore, the
awareness of gaze is beneficial to collaboration because collaborators
can use this communication modality to manage their interaction and to pinpoint
the possible interpretations of a referent. A strict relation between gaze and
collaborative work was demonstrated by Ishii and Kobayashi \cite{Ishii:1992fk}.
They showed that preserving the relative position of the participants
and their gaze direction could be beneficial for cooperative problem solving.
They used a system called \textsf{ClearBoard}, which allowed users to
collaboratively sketch on a shared display while maintaining eye-contact.
A similar setup was proposed by Monk and Gale \cite{Monk:2002lr}, which
they named \textsf{GAZE} system.

One of the limitations of the ClearBoard and GAZE prototypes was that of using half-silvered
mirrors to merge the remote image of the user with the computer display.
Unfortunately, users of these systems could see the reflection of their body
and hands on top of the remote image. During the experiments, this factor
emerged as bringing additional difficulties to the interaction. Additionally,
users were forced to interact in front of the camera because of the technology that was used to capture the video.
Therefore, their movements were constrained to the field of view of the cameras. Furthermore, using physical objects
in addition to the digital objects on the display was somewhat complicated
by the reflection and by the physical setups of the mirrors (see point
(b) of the introduction).

Simpler techniques used in the past to provide users with proactive control over
the focus of attention consisted in Media Spaces with multiple cameras. The
users of these systems could operate a manual switch to choose which camera
view was given to the remote user (see Gaver's {\em et al.}~\textsf{MTV} system
\cite{Gaver:1993oq}). Note that in a
face-to-face interaction, the control of the focus of attention is embodied and therefore
it does require a minimal effort. However, when using the manual selectors
in these systems users have to spend cognitive resources to keep track of
which view is offered to the remote collaborator (see point (b) of the introduction).
Furthermore, gaze awareness is naturally intertwined with
face-to-face communication whereas at a distance these two modalities have been decoupled in
many telepresence environments designed in the past (see point (a) of the introduction).

\subsection{Supporting Gestures and Deixis}

Gestures represent an extremely important communication mechanism that allows people to
coordinate their efforts and disambiguate their contributions in the interaction. In the last twenty
years, many solutions have been proposed to support remote gestures.
Many of the early prototypes used video technology to capture and display
the hands of the collaborators at the remote sites (\textit{e.g.}, \cite{Ishii:1992fk, Kirk:2007fk}). 
As Luff {\em et al.}~\cite{Luff:2003ij} clearly explained, video solutions
suffered from a fracture of the ecology of the remote sites. In such systems,
the gestures were fractured from the place where they were produced and where they were
received. Restricted field of views and distortion of projection are just few
examples of how video may hamper the usefulness of remote gestures.

An important limitation of these systems is due to the use of the unmediated video-capture
of the hands to communicate the gestures to the remote collaborator, as remote collaborators
had a hard time to infer when a particular gesture was associated to a communicative intent (see point
(b) of the introduction). Other research lead by Kuzuoka \cite{Kuzuoka:2004if}
proposed the use of robots on the remote sites to re-embody the interaction of the
remote collaborators. However, this research was conducted under the assumption
that one of the collaborators was the ``expert'' while the other was the ``novice''.
As highlighted in point (c) of the introduction, in face-to-face interactions these roles can
switch multiple times during the task resolution. Therefore, systems that are
designed around a static definition of the roles can introduce unnatural asymmetries in the
interaction.

Other researchers have proposed prototypes where gestures are represented
by digital metaphors like digitalized sketches or pointers. The underlying
assumption of this work was that a sketch could incorporate features of a
gesture that could suffice to replace the real gesture. More research is required
to define what features of gestures are most important for communication, what
kinds of gestures are necessary and in what circumstances.  For instance,
Gutwin and Greemberg \cite{gutwin2002} have reported mixed results in supporting workspace awareness for
collaborative work at a distance. One of their solutions included
a \textit{minimap} of the interface of the remote person that was shown as part of their system's interface.
This visualization reported in a schematic way the basic elements of the interface plus the information
of where the other was using his/her mouse pointer (\textit{i.e.}, \textit{telepointer}).
Other evaluations of the telepointer mechanism as a gesturing device have reported
negative results because the cursor activity could not be always related to
the user's intention, attention, or presence. Essentially, this type of tools present
some embodiment issues that lead to users' access and control disparities
(see point (b) of the introduction).

\section{Proposed Methodology}


In the last few years, computer vision techniques for real-time video processing have evolved quickly. We see
four major advancements that can help develop the new class of telepresence
environments: 

(1).~Several companies have released on the market \textit{eye-track-ing} solutions that
can trace the point of focus of a person's gaze moving freely in a room. Xuuk Inc.~recently
released a long-range eye sensor, called \textsf{eyebox2} which is able to detect
a user's gaze from up to $10$ meters of distance\footnote{{\scriptsize See \url{https://www.xuuk.com/}, last retrieved November 2009.}}. As discussed
in the previous section, a person's gaze can be mapped onto the focus of attention.
Hence, being able to automatically detect what the users are looking at is extremely
valuable to support interaction at a distance.

(2).~Real-time computer vision algorithms combined with sophisticated cameras enable the
detection and tracking of the hands and the \textit{recognition of the gestures} that a person is producing.
For instance, Miralles
{\em et al.}~\cite{Miralles:2009gd} have analyzed and tested a number of metaphors for the
definition of a gestural language to operate an interface as part of the Spanish-funded
\textsf{VISION project}. The camera used in their studies was originally produced by
3DV Systems Ltd\footnote{{\scriptsize See \url{http://www.3dvsystems.com/},
last retrieved November 2009.}}. Recently the company was bought by
Microsoft and the technology was
incorporated in the latest version of the \textsf{Xbox} gaming console.
The advantage of using this technology is that instead of
displaying the unmediated video of the hands to the remote site, the \textit{gist} of the
gestures that are mostly important can be captured and highlighted on the remote site
(\textit{i.e.}, gestures that are associated to communicative intents).

(3).~3D computer vision techniques allow to \textit{reconstruct the three-dimensional
volume of static and moving objects} in a certain scene from multiple camera-views \cite{Landabaso:2008ph}.
These techniques could be used
to infer the 3D position of the person in a certain environment (\textit{i.e.}, body tracking) and the
objects s/he is interacting with. This information could be combined with other sources
of information about the user's activity to build models of the user's actions and intentionality,
and to discern what to show or represent on the remote site.

(4).~Finally, advancements of vision techniques might allow to provide \textit{video devices
that can avoid the distortion of the gaze direction}. Within the EU-funded
\textsf{3DPresence} project a multi-perspective auto-stereoscopic 3D display has been developed
that is able to correct directional eye-contact and to render proper sense of perspective in
videoconferencing \cite{Divorra:2010aa}. Using this technology, conferees can feel simultaneously, and individually, whether they are being looked by other conference participants.

\begin{figure}[t]
\begin{center}
\leavevmode
\includegraphics[scale=0.62]{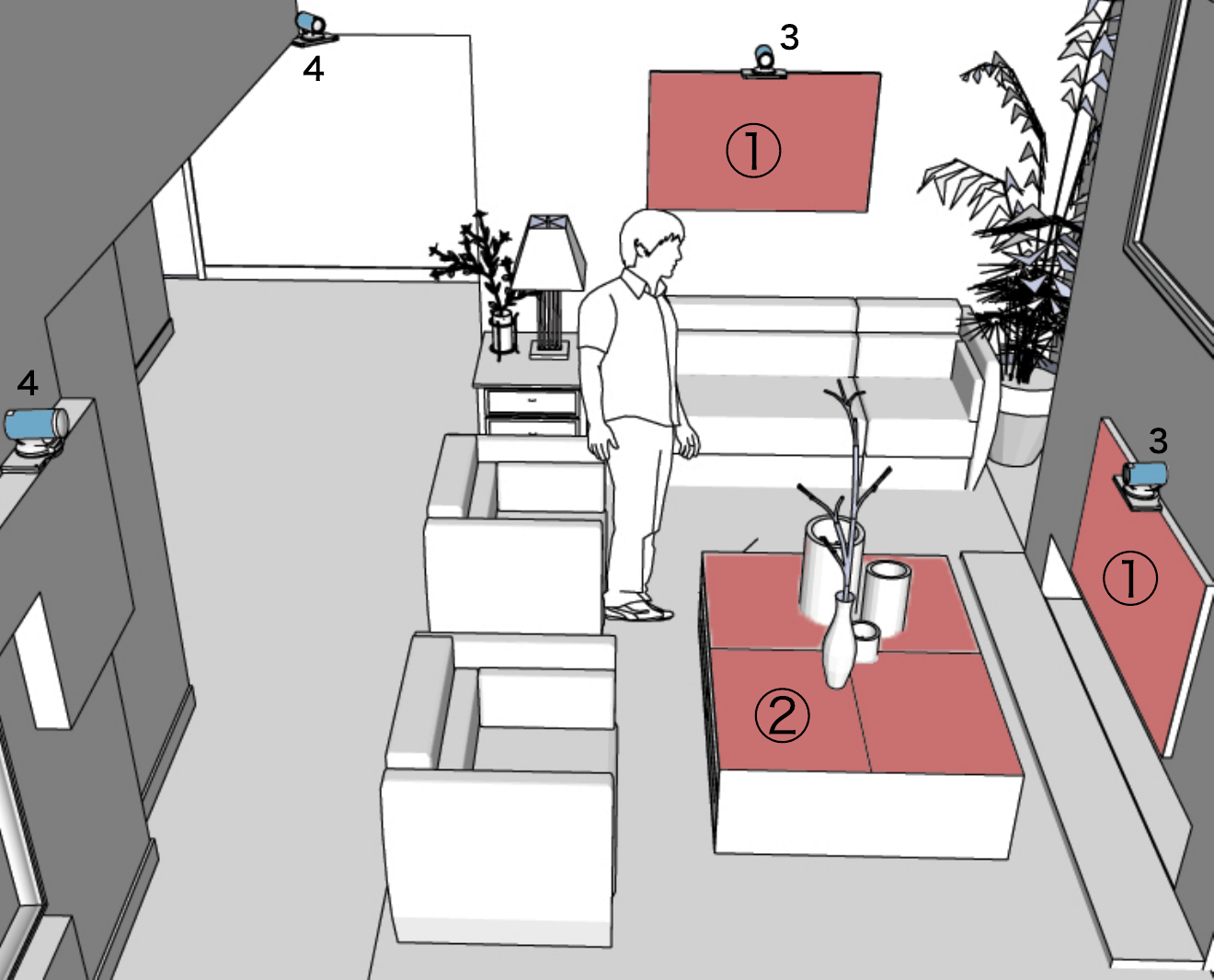}
\end{center}
\caption{{\small Possible experimental setup: (1) screens used to project remote information,
(2) shared workspace, 3 primary cameras used for eye-tracking, 4 additional cameras
used for 3D reconstruction of static and moving objects}}

\label{room_annotated}
\end{figure}

In summary, the three main points of our proposal are: (1) \textit{Multimodality}: We
intend to instrument an experimental environment where the three previously described technologies can
co-exist and complement each other to capture and model the user's activity (see
Figure \ref{room_annotated}); (2) \textit{Active Focus of Attention}:
Instead of limiting the communication to one modality,
the telepresence system should automatically detect and direct the remote
participant -- by means of the right communication modality --
to the part of the scene that is most relevant to the current situation.
Encouraging results in this sense have been obtained by Ranjan {\em et al.}~\cite{Ranjan:2007dk}; and (3) \textit{Evaluation}: We plan to test different combinations of communication
modalities, particularly analyzing the effects of these different mixes on efficiency
and ease of use. We initially plan to measure the efficiency
of the telepresence solution using task resolution time and we will measure ease
of use using the \textsf{NASA TLX} tool for measuring task load.

Our goal is to conduct a controlled experiment using a factorial design where we
manipulate the availability of \textit{non-verbal communication} --
\textsc{Gaze} and \textsc{Gesture} -- and the \textit{way this non-verbal
communication is transferred} to the remote site -- \textsc{unmediated}, when
the continuous feed of this information is provided and \textsc{gist} when the
information is processed to identify relevant episodes (\textit{e.g.}, one of
the participants indicates an object and says ``this''). Finally, we would like
to combine a factor related to the \textit{strategy used to show the focus of
the interaction} between the participants -- \textsc{manual} when it is left to
the user's choice, \textsc{automatic} when it is driven by the user's model, and
\textsc{semi-automatic} when it is set as a compromise between the two. Our
experiment will be similar to that of Monk and Gale \cite{Monk:2002lr} but
 with more sophisticated technology and combining more factors in the experimental
design.

\section{Expected Outcomes and Conclusions}
The experiment of Monk and Gale \cite{Monk:2002lr} showed that \textit{gaze awareness}
reduced the number of turns and number of words required to complete a task.
However, the setup they used forced the user to sit in a certain position
and at a certain distance from the half-mirrored screen. Also, subjects could see all
the time the position of the eyes of the other participants, even when gaze was not
associated to communicative intent. We believe that redesigning this experiment
with modern technology might provide concluding evidences on the usefulness of
gaze awareness on collaborative work at a distance.

Furthermore, we believe that further research is necessary to understand how gaze
awareness should be combined with \textit{gesture recognition} and the effect of
the availability of both modalities on remote collaboration. While it is clear that support for
gaze awareness and gestures is important in CSCW-environments,
there is evidence that providing this information continuously might
be detrimental to problem-solving, as it increases the amount of effort required to the participants.
We believe that the experiments proposed in this paper will yield relevant implications
in the design of mechanisms that could \textit{mediate} the representation of both
non-verbal communication modalities.

For instance, Cherubini {\em et al.}~\cite{Cherubini:2008uq} demonstrated
that the probability of misunderstandings between
distant collaborators in problem solving task is related to the distance
between the collaborators' focus of gaze over the shared workspace. Therefore,
Cherubini's results support a telepresence solution where the focus of gaze is shown to the
remote site {\em only} when the system infers that there might be a misunderstanding
between the collaborators. The experiment we are proposing here will
compare this mediated modality of representing gaze and gesture to the
remote site with the unmediated approach.

To conclude, this paper briefly describes some of the telepresence ideas we plan to
work on in the near future. We intend to contribute to the workshop with these
concepts and to receive relevant feedback on our proposed approach.
Finally, we hope to stimulate rich discussions on the future of telepresence.

\section{Acknowledgments}
Telef\'{o}nica I+D participates in Torres Quevedo subprogram (MICINN), 
cofinanced by the European Social Fund, for Researchers recruitment.

\bibliographystyle{abbrv}

\end{document}